# Quantifying Spatiotemporal Stability by means of Entropy: Approach and Motivations

Telecom Sud Paris Research Report #RS2M10001

Mohamed-Haykel Zayani
RS2M Department
TELECOM & Management SudParis
Évry, France
mohamed-haykel.zayani@it-sudparis.eu

Vincent Gauthier
RS2M Department
TELECOM & Management SudParis
Évry, France
vincent.gauthier@it-sudparis.eu

Djamal Zeghlache
RS2M Department
TELECOM & Management SudParis
Évry, France
djamal.zeghlache@it-sudparis.eu

*Abstract*—Several studies demonstrate that there are critical differences between real wireless networks and simulation models. This finding has permitted to extract spatial and temporal properties for links and to provide efficient methods as biased link sampling to guarantee efficient routing structure. Other works have focused on computing metrics to improve routing, specially the reuse of the measure of entropy. From there, rises the idea of formulating a new measure of entropy that gives an overview of the spatiotemporal stability of a link. This measure will rely on spatial and temporal properties of links and fed with the efficiency of biased link sampling.

## I. INTRODUCTION

Wireless networks and particularly wireless sensor networks have been closely studied in last years. Many works [1, 2, 3, 4] have shown that the wireless communication networks have wide and deep differences with simulation models. Recent studies [5, 6, 7, 8] have attempted to make the light on characteristics of these networks.

Cerpa et al. [5] have presented interesting temporal properties of low power wireless links relying on experimentations and have proposed from them a new centralized routing algorithms and a distributed one. Zuniga and Krishnamachari [7, 8] have tried to explain the notion of "transitional region", giving some spatial properties of the same type of links and providing mathematic analysis to understand the properties of this region and its behavior.

Other ones have targeted to show the efficiency of sampling to find a satisfactory routing structure. Zhang et al [10, 11] have proven the efficiency of data-driven link estimation, using biased link sampling, to converge to an efficient routing solution, even with dynamic environment parameters (changing traffic, different topologies).

It is also important to note that several work aim to propose new routing solutions [15, 21, 22, 23, 24, 25]. Among the different propositions, the measure of entropy seems as relevant one. The entropy is inspired from thermodynamics and deducted from the analogy between nodes in a network and mixtures of gazes. It provides a quantification of the order or the stability of a situation.

In this paper, we will expose a detailed summary of the works cited above and try to extract perspectives to future works. Our aim is to provide a new entropy measure formulation that will be able to quantify the degree of spatiotemporal stability of a link.

The rest of the paper is organized as follows. In section II, we provide a state of the art of the major work that related to spatial and temporal properties wireless networks links and underline the importance of biased link sampling to converge to good solution for routing. We will present, in the section III, the mathematical analysis of Zuniga and Krishnamachari [7, 8] for the properties of the "transitional region" and give some other results that have aroused our curiosity. In the same section, the metrics used in [5] and [11] are mentioned. The section IV focuses on the measure of entropy; its origins and using in different domains and particularly in the domain of wireless networks. The main details of our future work are developed in the section V and concluding is made in section VI.

## II. RELATED WORK

Several studies have focused on the comparison between empirical observations of low power wireless communication networks and frequently used simulation models [1, 2, 3, 4]. These studies have shown that the latter ones do not reproduce faithfully the real aspects of wireless networks and lead to incoherent results. Nevertheless, they have not analyzed sufficiently the temporal characteristics of wireless links.

Cerpa and al. [5] study statistic temporal properties of links in the low power wireless communication networks. Hence, different experimental scenarios, using the SCALE wireless measuring tool [4] were considered to understand some network short-term and long-term behaviours (single link auto-correlation, bandwidth quality, covariance between links with same source, correlation between forward and reverse links, effects of different packets sizes, links temporal consistency, correlation between path links). The experimentations have contributed to the extraction of a set of relevant properties permitting to insure more efficiency in routing protocols. Indeed, a strong temporal correlation implies the use of required number of packets for reception as quality metric instead of reception rate. This correlation stipulates also using only good links. Moreover, the variance in time lagged correlation between forward and reverse links express the necessity to send acknowledgements immediately after the reception. Meanwhile, a lower variance promotes the use of longer packets. Given these results and considering the correlation between links of a same path and the consistency of good links, two new shortest path algorithms are proposed. In the one hand, a centralized algorithm which is a generalization of Dijkstra algorithm that takes into consideration the correlation between successive links of a path. In the other hand, a localized probabilistic algorithm that affects probabilistic gradients to forward links with referring to statistics of reverse links.

As Cerpa et al. [5], Woo et al. [3] and Zhou et al. [6] have proposed link models relying on experimental conclusions. Nonetheless, they don't give mathematical approaches to show the effects of channel and radio dynamics on links unreliability and asymmetry. Moreover, Cerpa et al. [5] and Woo et al. [3] present the inconvenient of being closely tied to specific channel and radio parameters adopted in experimentations.

Zuniga and Krishnamachari analyze the major causes behind unreliability [7, 8] and asymmetry [8] of low power wireless sensor networks. In this kind of network, the use of binary disc-shaped model to model the range is inappropriate. Indeed, it exists a "transitional region" [1, 2, 3] which is unpredictable toward the good reception of a packet and affects the upper-layer protocols reliability. To understand this region, two models have been proposed: a channel model that is based on the log-normal path loss propagation model [9] and a radio reception model closely tied to the determination of packet reception ratio. The work on these models introduce to the formulation of the expressions of distribution, expectation and variance of the packet reception ratio according to the distance. These approaches have contributed to the determination, in terms of distances (of the beginning and the end) of the transitional region and the computation of its coefficient (relative to its size). The coefficient is more important with lower path loss exponent and higher deviations. It has been also proved that this region remains even though the presence of perfect-threshold receiver because of multipath effects. For the question on the asymmetry, heterogeneous hardware contributes to the extension of the transitional region. Nevertheless, a negative correlation between output power and noise floor decreases the effects of asymmetry levels.

Thus, wireless communication is described by two complex sides: the spatial side and the temporal side. The estimation of link properties stands out as an essential recourse in routing. The data-driven link estimation [10] is the best method to do it; the information about the properties of the link is collected by the MAC feedback with unicast data transmission along the link. If the link is not used, the properties will be considered as unknown. This brings us to the biased link sampling (BLS) issue where properties of active links are sampled and updated whereas the inactive ones are not sampled and unknown.

Zhang et al., in [10], via various testbeds of 802.11b networks, aim to demonstrate that beacon-based link estimation suffers from several drawbacks (impact of environment, packet type, packet size, interferences…). They also show that it is difficult to have a precise estimation, even with length and transmission rate as those of data packets. To address this problem, they propose to estimate link properties via MAC feedback collected from unicast data transmissions, and precisely the MAC latency. In this way, they define data driven routing metrics (ELD metric, as most important one, which represents a data driven version of ETX/ETT metric (ETX for expected number of transmission required to successfully deliver a unicast packet, ETT for expected time of transmission required). Building on this capability, Zhang et al. design a routing protocol "Learn On the Fly" (LOF) where estimation relays on unicast data transmissions. The proposed protocol uses control packets only during booting up; each node discovers its neighbours by taking few samples on the MAC latency. Then, the node adapts its routing decisions considering only the MAC feedback. The protocol LOF presents also the characteristic of taking into consideration temporal variations and eventual imperfection of initial estimations; each node explores alternatives neighbours with a certain probability and controlled frequency. The collected observations confirm the contribution of LOF in reducing MAC latency and energy consumption in packet delivery while improving route stability and network throughput (in the cases of bursty event and periodic traffics).

In [11], Zhang et al. focus on the effects of BLS on the convergence of routing to optimal solution, using mathematical analysis and testbed experimentations for wide-range of scenarios (grid or random topology, diverse traffic levels). They also use the ETX metric (expected number of transmission required to successfully deliver a unicast packet) to identify the best forwarder of the routed packet. Their works prove that the choice of the next forwarder in a routing structure remains unchanged, even though there are important changes of the network condition and the traffic intensity. In the case where optimal routing structure changes, after the worsening of network conditions, the reaching of the new optimal solution is guaranteed. The convergence presents an interesting characteristic of being quick. A small number of unicast sampling packets is sufficient (no more than 7). When the network conditions are better, the convergence to an optimal solution is not assured (due to BLS properties). However, the routing structure, which was able to support a heavy traffic, can handle a lighter one. This structure is a sub-optimal is reliable and has performances very close to the optimal solution.

III. USED TOOLS TO MEASURE METRICS

The major works described below, in the related works, have measured some interesting metrics to provide results and analysis. Zuniga and Krishnamachari [7, 8] propose the computation of the packet reception rate to understand the causes behind the move and the extent of the transitional region. For this, they present a mathematical model that we present briefly in the following subsection. We will also have a look on the empirical measurements done in the works of Cerpa et al. [5] and Zhang et al. [11].

## A. Packet Reception Rate (PRR)

### 1) Mathematical modelisation

Zuniga and Krishnamachari [7, 8] use mathematical techniques from communication theory to model and analyze these links. The main objective is to identify the causes of the transitional region and the quantification of their influence. For this aim, expressions for the packet reception rate as a function of distance will be derived. These expressions take into consideration radio and channel parameters such as the path loss exponent, the channel shadowing variance, the modulation and encoding of the radio. The model does not consider interference and rely on the assumption that scenarios consider the traffic and contention very light.

The approach followed by Zuniga and Krishnamachari tries to show how the channel and the radio determine the transitional region. On the one hand, for the wireless channel, the log-normal shadowing path loss [9] model is adopted (can be used for small and large coverage systems and its accuracy is demonstrated in comparison with other models). It is given by:

$$PL(d) = PL(d_0) + 10\eta \log_{10}\left(\frac{d}{d_0}\right) + N(0,\sigma) \quad (1)$$

Where $d$ is the transmitter-receiver distance, $d_0$ is a reference distance, $PL(d_0)$ is the power decay for the reference distance $d_0$, $\eta$ is the path loss exponent and $N(0, \sigma)$ is a zero-mean Gaussian random variable with standard deviation $\sigma$. Hence, for an output power $P_t$, the received power $P_r$ in dB is expressed by:

$$P_r(d) = P_t - PL(d_0) - 10\eta \log_{10}\left(\frac{d}{d_0}\right) + N(0,\sigma) \quad (2)$$

On the other hand, for the radio, the packet reception rate $\Psi$ of successfully receiving of a packet, using a modulation M is:

$$\Psi(\gamma) = (1 - \beta_M(\gamma))^f \quad (3)$$

Where $\gamma$ is the SNR (Signal-to-Noise Ratio), $\beta_M$ is the bit-error rate and a function of the SNR and $f$ is the frame size. The SNR at a distance $d$ can be expressed from (2):

$$\gamma(d)_{dB} = P_r(d) - P_n = N(\mu(d),\sigma) \quad (4)$$

Where $N(\mu(d), \sigma)$ is a Gaussian random variable with mean $\mu(d)$, variance $\sigma^2$ and $P_n$ is the noise floor. Moreover, the expression of $\mu(d)$ can be determined from (2) into (4):

$$\mu(d) = P_t - PL(d_0) - 10\eta \log_{10}\left(\frac{d}{d_0}\right) - P_n \quad (5)$$

Denoting the bit-error rate for the SNR in dB as $BER_M(\gamma_{dB}) = \beta_M(10^{\gamma_{dB}/10})$, the packet reception rate $\Psi$ can be expressed by:

$$\Psi(\gamma_{dB}) = (1 - B_M(\gamma_{dB}))^f \quad (6)$$

In [8], Zuniga and Krishnamachari underline the impact of hardware variance, they propose the expression of the SNR measured at a node B for the output power of a node A and denoted as $SNR_{AB}$ ($\gamma_{AB}$):

$$\begin{aligned}\gamma_{AB} &= P_{tA} - PL(d) - P_{nB}\\ &= N(P_t, \sigma_{tx}) - PL(d) - N(P_n, \sigma_{rx}) \quad (7)\end{aligned}$$

Where $\sigma_{tx}^2$ and $\sigma_{rx}^2$ represent the variance of the output power and the noise floor respectively.

Empirical results in [8] show that it exists correlation between output power and noise floor. These two latter powers are taken as multivariate Gaussian distribution:

$$\begin{pmatrix} T \\ R \end{pmatrix} \cong N\left(\begin{pmatrix} P_t \\ P_n \end{pmatrix}, \begin{pmatrix} S_T & S_{TR} \\ S_{RT} & S_R \end{pmatrix}\right) \quad (8)$$

Where $P_t$ is the nominal output power, $P_n$ is the average noise floor, $T$ and $R$ are the actual output power and noise floor, respectively, of a particular radio and $S$ is the covariance matrix between the output power and the noise floor. From the experiments in [8], the matrix $S$ is given by:

$$S = \begin{pmatrix} 6.0 & -3.3 \\ -3.3 & 3.7 \end{pmatrix} \quad (9)$$

The randomness of the SNR in (4) is provided only by multipath effects. The variances of the output power and the noise floor add two other factors of randomness. The new expression of SNR becomes:

$$\begin{aligned}\gamma &= N(P_t, \sigma_{tx}) - PL(d) - N(P_n, \sigma_{rx}) \\ &= N(P_t - P_n, \sigma_{hw}) - PL(d) \\ &= N(P_t - P_n, \sigma_{hw}) - \overline{PL(d_0)} + N(0, \sigma_{ch}) \\ &= N(P_t - \overline{PL(d_0)} - P_n, \sigma_t) \quad (10)\end{aligned}$$

Where $\sigma_{hw}^2 = \sigma_{tx}^2 + \sigma_{rx}^2$, $\sigma_t^2 = \sigma_{hw}^2 + \sigma_{ch}^2$ and

$$\overline{PL(d_0)} = PL(d_0) + 10\eta \log_{10}\left(\frac{d}{d_0}\right)$$

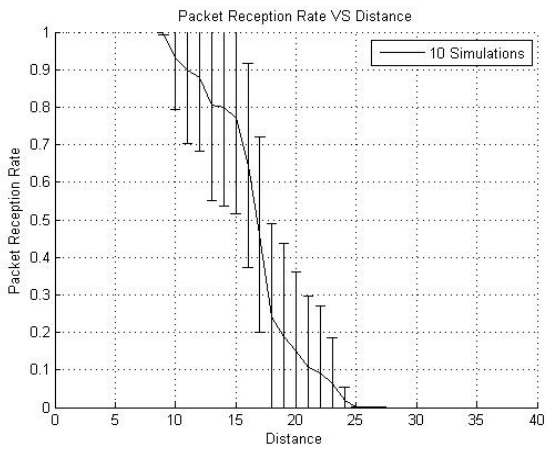

(a) 10 Simulations

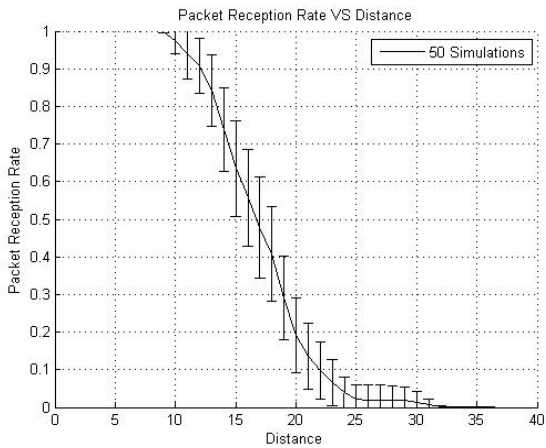

(b) 50 Simulations

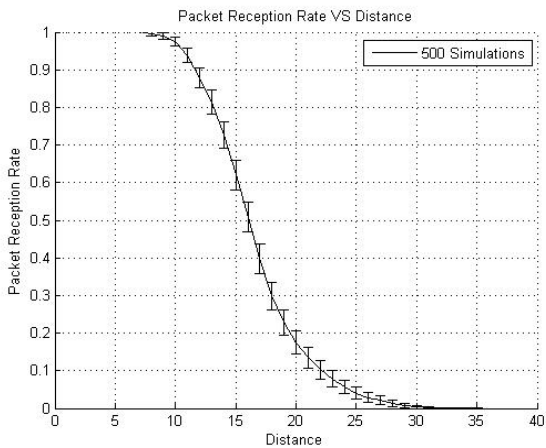

(c) 500 Simulations

Figure 1.  Packet Reception Rate VS Distance for different simulations numbers

*2) Experimental tests*
  *a) Experimentations on different numbers of simulations*

We propose to display average packet reception rate as function of time in the aim to see the variations of confidence intervals for different numbers of simulations. We reuse the same parameters as in [7] with NCFSK (Non Coherent Frequency Shift Keying) Modulation and Manchester encoding. The objective is to see the behaviour of confidence intervals and standard deviations of average packet reception rate with different number of simulations.

Figure 1 shows the variation of average packet reception rate with the distance for different simulations numbers. We remark that the higher the simulations number the smaller the confidence intervals. For each case, the confidence intervals are larger in the transitional region (0.2<PRR<0.8) because of packet reception rate unpredictability in this region. A high number of simulations enable us to extract finest confidence intervals. When the number is low, the information is not sharp as below.

Figure 2 represents the evolution of average packet reception rate standard deviation as function of distance. We see that the standard deviations for each scenario take maximum values (and almost the same for different cases) when the distance varies into the transitional region (in the same way with Figure 1). Nevertheless, the Gaussian form is larger with higher number of simulations. We can explain this by the fact that there are more chances to have random values of PRR as we are increasingly close to the frontiers of transitional region.

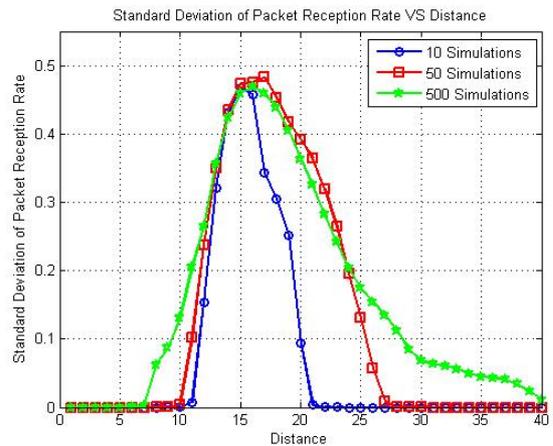

Figure 2.  Average Packet Reception Rate Standard Deviations VS Distance for different simulations numbers

  *b) Experimentations on different numbers of nodes*

We also try to see how behave confidence intervals and standard deviations when we use one node for 1000 simulations and 10 node for 100 simulations each one. Will the second scenario cause more randomness?

For this, we take into consideration that there are randomness caused by multipath effects and hardware variance as in [8]. We use also parameters used in [8] to work on indoor and outdoor environments. We keep the use of NCFSK as modulation technique and Manchester as encoding scheme.

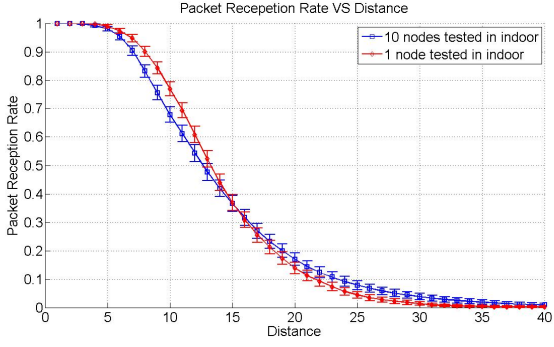

(a) indoor

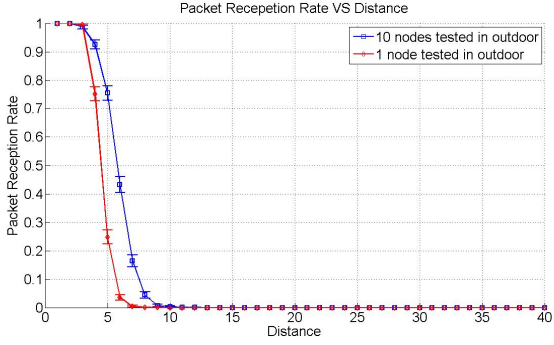

(b) outdoor

Figure 3. Average Packet Reception Rate VS Distance for different sets of nodes

Figure 3 shows the evolution of average packet reception rate with the distance for 1000 simulations done with one node and with 10 nodes. We find that the average PRR for the two scenarios, in both environments, have close evolutions. We notice also that in the transitional region, the confidence intervals have almost the same sizes. At the borders of the region, the confidence intervals in the scenario with 10 nodes are a little bit larger. This demonstrates that if we are far from borders of transitional region, the different sources of randomness are "hidden" by the unpredictability of the region. When we are closer to the borders, the effects of randomness sources are visible because we are nearer to predictable region (connected or disconnected).

Figure 4 represents the evolution of packet reception rate standard deviation with the distance for the same scenarios. We detect some similarities with Figure 2. When we have several nodes, there is a combination between many configurations of randomness. Using one node limits the randomness. That's why, when 10 nodes are used, the Gaussian evolution is larger.

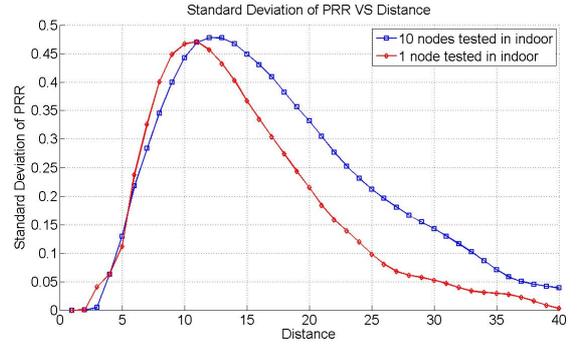

(a) indoor

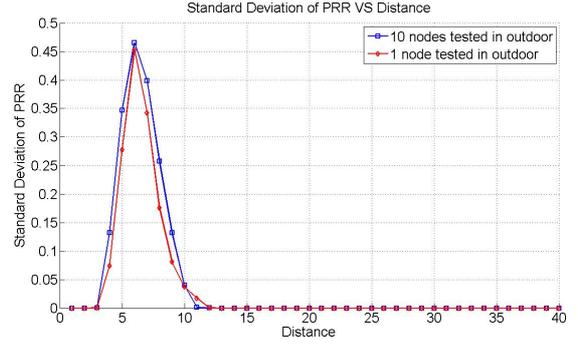

(b) outdoor

Figure 4. Packet Reception Rate Standard Deviation VS Distance for different sets of nodes

### B. Data-Driven Versions of ETX/ETT Metric

*1) Mathematical modelisation*

Zhang et al. [11] have used in their study the ETX metric (expected number of transmissions for delivering a data packet). To estimate it, they consider the data-driven link estimation and routing method L-ETX [10]. In the latter method, the packet delivery reliability (PDR), along a link, is determined from MAC feedback for unicast data transmissions. The ETX of this link is computed as the inverse of the PDR. The measure for a path consists in summing the ETX metrics of each link belonging to the path.

Zhang et al. have also considered a localized, geographic routing metric ETD (ETX per unit-distance to destination) to have an idea on the efficiency of forwarding neighbours. Taking into consideration $S$ as sender, $R$ as neighbour of $S$ and $D$ as destination, the ETD via $R$ is given by

$$\begin{cases} \dfrac{ETX_{S,R}}{L_{S,D} - L_{R,D}} & \text{if } L_{S,R} > L_{R,D} \\ \infty & \text{otherwise} \end{cases} \quad (11)$$

Where $ETX_{S,R}$ is the ETX of the link from $S$ to $R$, $L_{S,D}$ represents the distance that separates $S$ and $D$ and $L_{R,D}$ the distance between $R$ and $D$.

In [10], a similar approach is adopted. The MAC latency appears as a routing metric similar to ETT (and ETX with fixed transmission rate). From this consideration, the ELD metric is defined (ELD for expected MAC latency per unit distance to destination) which permits to minimize the end-to-end MAC latency from source to destination.

The MAC latency per unit distance to the destination (LD), denoted by *LD(S, R)*, is given by

$$\begin{cases} \dfrac{D_{S,R}}{L_{S,D} - L_{R,D}} & \text{if } L_{S,R} > L_{R,D} \\ \infty & \text{otherwise} \end{cases} \quad (12)$$

Where $D_{S,R}$ is the MAC latency from S to R. The expected MAC latency per unit distance to destination ELD, denoted by ELD(S, R), is calculated as following

$$\begin{cases} \dfrac{E(D_{S,R})}{L_{S,D} - L_{R,D}} & \text{if } L_{S,R} > L_{R,D} \\ \infty & \text{otherwise} \end{cases} \quad (13)$$

A source will choose the next hop that gives the lowest ELD metric among all the neighbours. This recourse helps to select forwarders, in a reliable communication range, to decrease the end-to-end MAC latency and in the same time to reduce energy consumption.

Cerpa et al. [5] used also the ETX metric that they call RNP (Required Number of Packets) and the packet reception rate. They compute them empirically with the set of tests that they propose. They also underline that assuming that ETX metric is the inverse of the packet reception rate is not an efficient estimation when the latter is between 10% and 90%.

*2) Experimentations*

We propose to represent the ETD metric jointly with the PRR evolution. Taking the indoor configuration, we assume the following scenario: we consider a sender S, a destination D and a relay node R. We vary the distance between the node S and the node R and compute for each distance the average ETX metric (with confidence intervals) and then the ETD metric. The distance between the node S and D is 40 meters and the distance between S and R varies between 1 and 39 meters. We assume also that the maximum number of retransmissions is 7. When the transmission is failed, ETX is equal to 8. We represent the results for 100 and 200 simulations.

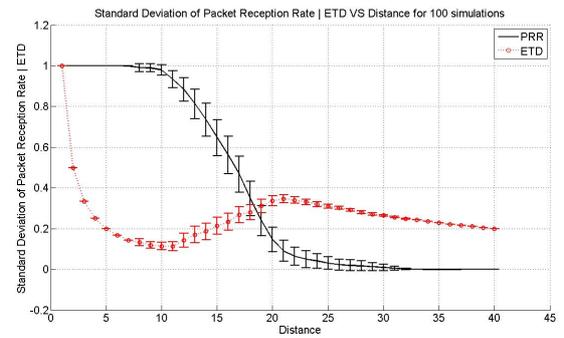

(a) 100 simulations

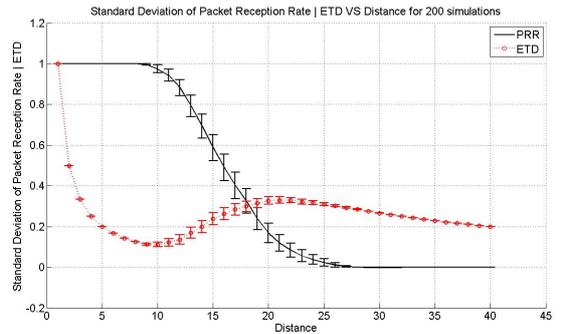

(b) 200 simulations

Figure 5. Average Packet Reception Rate and ETD metric VS Distance for different simulations number

The results are given in the figure 5. In the connected area, ETD is lower with more important distance between the node S and the node R. This result is logic because in the connected area the number of retransmissions is always 1. We note that the lowest value of ETD corresponds to a distance that marks the beginning of the transitional region (11 meters with 100 simulations and 10 meters with 200 simulations). At this distance, the PRR (0.933 with 100 simulations and 0.974 with 200 simulations) is still important and the losses are few. This observation indicates that the relay node far by this distance from the node S, is the best forwarder because it permits to be as nearest as possible to the destination and keeping in the same time a good delivery rate with the sender. We see that the ETD metric takes higher values when the relay node is in the middle of transitional region; this is due to the increase of the number of retransmissions. When the relay node is near to disconnected area, ETD metric decreases linearly; the average number of retransmissions tends to 8 (as we assume when the transmission is failed, we note also that confidence intervals are small) and with incremental distances, the ETD metric is smaller.

IV. THE MEASURE OF ENTROPY

The measure of entropy has been used in several works. Its definition has varied from one work to another. We will expose a little overview on the origin of entropy and a brief description of works having used this measure.

*A. Original definition of Entropy*

The concept of entropy was introduced for the first time by Clausius [12] as a unique measure of reversible change in thermal energy concerning the absolute temperature. He focused on the macroscopic behaviour of chemical microscopic reactions and proposed thermodynamic entropy. Based on Clausius works, Boltzmann defined the combination microstates statistic entropy [13] as

$$S = -k_B \sum_i p_i \ln p_i \quad (14)$$

Where $p_i$ is the probability that the microstate $i$ is verified during system fluctuations and $k_B$ is Boltzmann's constant. This definition is applied to characterize the order in the system and how the system self-organizes among different entities.

Later, Shannon introduced the concept of information entropy H [14]. This measure has been used to quantify the capacity of a transmission channel and has been extended to other domains.

*B. Different approaches using entropy*

Many works propose the use of entropy, for different aims. In [15], Lu et al. have presented, firstly, the principles of self-organization. Indeed, wireless networks use the self-organization to minimize configuration needs, to facilitate the deployment of the network and to support applications and services. The recourse to self-organization schemes permits to improve the order in the network. This organization take place on two levels: microscopic (logic links between nodes) and macroscopic (formation of flexible structure). Secondly, Lu et al. justify the use of entropy with three major reasons:

- The organization in a wireless network is similar to a thermal dynamic system.
- Many metrics have been proposed to evaluate self-organization strategies (protocol overhead, algorithmic complexity…) but they do not give any idea on order degree.
- The statistic entropy used in thermodynamics is a key measure because it describes the behaviour of self-organization protocols compared to changes of inherent parameters in the network as the reliability of links and nodes.

Lu et al. consider the entropy as a system macroscopic description taking into consideration microscopic interactions. Similarly, the equilibrium between two perfect gazes (macroscopic level) is the result of molecular interactions (microscopic level). Since self-organization limits interactions, so it limits also the entropy. To evaluate the entropy of a link, the following expression is proposed

$$E = -\sum_{u,v \in X} p(u,v) \log_2 p(u,v) \quad (15)$$

Where $u$ and $v$ represent two nodes from the network nodes set $X$.

High entropy values indicate an important disorder, whereas low ones signify a better organization. This analysis is inspired from thermodynamics: when the entropy is low, the equilibrium is more stable and the disorder is less significant at molecular level.

This approach is interesting but the measure of entropy give only an overview of the order quantification and do not inform about stability and robustness of routing when used with self-organization strategies.

In [16], Sneppen et al. use measures applied on network topologies. These measures characterize the ability of a node to lead and send a signal to its destination(s). The entropy is considered and is described as the capacity to predict from which neighbour the message arrives. It quantifies the predictability (or the order/disorder) of a traffic around a node. The probabilities represent the fraction of messages received from each neighbour. In [17], they present measures the investigation on constraints posed by the network structure on communication. They define two measures of entropy: the predictability relative to messages targeting a specified node (Target Entropy) and the predictability relative to messages crossing a specified node and one of its neighbours. The analysis of theses measures lead to some conclusions:

- When the entropy values are high, the predictability is low. In the opposite case, a little number of links is used.
- The traffic to nodes with high degrees is unpredictable.
- Low values of entropy show that the traffic is concentrated. It is more distributed and logically more robust when the entropy is higher.

In [18], Van Dyke Parunak and Brueckner try to show that the relationship between self-organization in multi-agent systems and the second law of thermodynamics is not a metaphor and that this relation can provide analytic and quantitative directives in the aim of the conception and the deployment of these systems. The self-organizing model is inspired from [19], which suggests that the key idea permitting to reduce the disorder in a multi-agents system is to copulate it with an another where the disorder increase. For a system, the self-organization is done at macroscopic level. Considering this, such behaviour contradicts the second law of thermodynamics. Nevertheless, the system includes a microscopic level which dynamic increase the disorder. To reproduce a system with the two levels, the example of the pheromones used by ants is considered. The movements of ants constitute the macroscopic level while the molecules of pheromones represent the microscopic level. The movements of ants permit to define a little number of ways between the nest and the source of food. The disorder at macroscopic level is inconspicuous. The latter observation is the result of coupling macroscopic level agents with microscopic level, where the evaporation of molecules of pheromones takes place according to random mobility which increases the disorder.

The entropy, inspired from Shannon entropy and thermodynamic entropy, defines a disorder measure describing the trend of a system to be chaotic. This measure is applied at two levels: localisation (microscopic level) and direction (macroscopic level). The results show that the two measures are antagonist which confirms that the second law of thermodynamics can be applied to multi-agent systems.

Another approach [20] tries to provide a schema able to classify a connection by three categories: Ethernet, WLAN or connexion with low bandwidth. An algorithm is proposed to identify the connection type using the sending of packets pairs. This choice is led by the motivation to follow the random aspect at the reception of the pairs and use this aspect to identify the type of connection. The measure of random aspect is done by Shannon entropy. The major reason of this recourse instead the use of variance is that entropy is a better metric catching the random aspect of a random variable.

As final note, it is interesting to mention that some works have used entropy to determine path stability in MANET and wireless sensor networks for their respective routing protocols. EQMGA (Entropy-based model to support QoS Multicast routing Genetic Algorithm) [21], ERPM (Entropy-based Routing Protocol using Mobility) [22], the An and Papavassiliou model [23], ELMR (Entropy-based Long-life Multipath Routing algorithm) [24] and QARPE (QoS-Aware Routing Protocol based on Entropy) [25] share the same idea of constructing a new entropy and select the most stable path relying on the entropy to reduce the number of route reconstruction when the topology is continuously changing.

To understand these approaches, we expose the reasoning followed by them. Indeed, to a node $m$ is associated a set of variable features $a_{m,n}$, where node $n$ is a neighbour of node $m$. These variable features express the relative speed among two nodes. With changing networks, changes of $a_{m,n}$ values are expected. Considering $v(m,t)$ and $v(n,t)$ respectively the velocity vectors (expressed by direction and speed) of nodes $m$ and $n$, the relative velocity between the two nodes at time $t$ is:

$$v(m,n,t) = v(m,t) - v(n,t) \quad (16)$$

In [22] and [23], an expression of variable features $a_{m,n}$ is proposed according to the relative speed:

$$a_{m,n} = \frac{1}{N}\sum_{i=1}^{N}|v(m,n,t_i)| \quad (17)$$

Where $N$ is the number of discrete times $t_i$ that velocity information can be calculated and disseminated to other neighbouring nodes within time interval $\Delta_t$.

However, in [21], [24] and [25], $a_{m,n}$ variables take also the relative position $p(m,n,t)$ expressed by:

$$p(m,n,t) = p(m,t) - p(n,t) \quad (18)$$

Where $p(m,t)$ and $p(n,t)$ are respectively the position vector of nodes $m$ and $n$ at time $t$.

The expression of $a_{m,n}$ variables is given by:

$$a_{m,n} = \frac{1}{NR}\sum_{i=1}^{N}\left[\begin{array}{c}|p(m,n,t_i) + v(m,n,t_i)\Delta_{t_i}| \\ -|p(m,n,t_{i+1})|\end{array}\right] \quad (19)$$

Where $R$ is the radio range.

From this, the definition of entropy $H_m(t,\Delta_t)$ at node m during time interval $\Delta_t$ is expressed as follows:

$$H_m(t,\Delta_t) = \frac{-\sum_{k \in F_m} P_k(t,\Delta_t)\log P_k(t,\Delta_t)}{\log C(F_m)} \quad (20)$$

Where $P_k(t,\Delta_t) = \dfrac{a_{m,k}}{\sum_{i \in F_m} a_{m,i}}$

In this expression, $F_m$ denotes the set or any subset of neighbors of node m. The parameter $C(F_m)$ is the cardinality of the set $F_m$. It is clear that $H_m(t,\Delta_t)$ is normalized so its values vary between 0 and 1. Low values of the entropy express that the change of the variable values is important. The contrary signifies that this change is limited.

With this finding, some measures of route stability between two nodes have been proposed:

$$RS_1 = \prod_{i=1}^{N_r} H_i(t,\Delta_t) \quad (21)$$

$$RS_2 = -\ln RS_1 = -\sum_{i=1}^{N_r}\ln H_i(t,\Delta_t) \quad (22)$$

V. PERSPECTIVES

Our objective in the future works is to exploit the results exposed in the related work and to propose the entropy as stability indicator. In the one hand, the works of Cerpa et al. [5] and, Zuniga and Krishnamachari [7, 8] give to us interesting spatial and temporal properties. On the other hand, we will try to use entropy to guess the degree of stability in the spatial side (stability of the neighbourhood, mobility …) and in the temporal side (stability of measures, neighbourhood…).

The objective is to present an entropy measure that gives an idea on spatiotemporal stability of a node in the network. These measures will be done as by the biased link sampling (BLS) viewed in [11] or as in LOF [10].

The manner that the measure of entropy would be computed will be defined later. Nevertheless, it will follow the reasoning below. We suppose that we have a network and we want to have an idea on the stability of a node and the efficiency of its links.

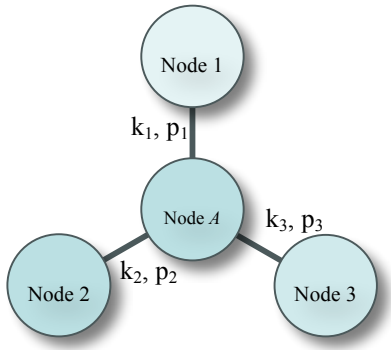

Figure 6. Example of a node and its neighbours, with spatiotemporal stability (p) and quality of links (k) parameters

In Figure 6, we present a little example of a node $A$, in a network, with three neighbours. The node $A$ keeps two types of measures:

- The spatiotemporal stability $p_i$ with the neighbour $i$.
- The link quality $k_i$ of the link between the node $A$ and the node $i$.

We propose, using these parameters, to compute a measure of entropy that gives an idea on the efficiency of a link after dynamic changes (spatial and temporal). A possible measure can be formulated as following:

$$E = \frac{1}{N} \sum_{i \in N_A} k_i p_i \log p_i \quad (23)$$

Where $N_A$ is the set of neighbours of the node $A$ and $N$ is the number of neighbours of node $A$.

We can take the formulation of entropy, proposed in [21, 22, 23, 24, 25] as starting formulation and we improve it. In our mind, it is essential, for a node, to consider neighbors mobility, but it is not sufficient. There are other parameters that must be taken into consideration, as link efficiency towards interferences and noises or the presence of the neighbor in the transitional region. The latter scenarios complicate the interactions between them.

After the definition of this new formulation, we will do testbed experimentations to gather results in order to validate the efficiency of our proposal. We intend using a recent mobility model called TVC model [26]. It is presented as the first synthetic mobility model able to capture non-homogeneous behavior jointly in space and time.

The TVC model is constructed from WLAN traces and is characterized by three major contributions. First, it is the first model able to capture time-variant mobility characteristics (location visiting preferences and time-dependent mobility behavior). Second, it can recreate mobility behavior from the traces and is mathematically tractable. This is very useful for prediction in order to evaluate protocols performances (mathematic models that capture average node degree, hitting time and meeting time). Finally, the TVC model is not reduced to some scenarios but matches with qualitatively different traces (different WLAN traces, vehicular traces, human encounters traces).

We also want to incorporate a model that reproduces the impact of interferences (since that Zuniga and Krishnamachari model does not take it into consideration). We focus on Qiu et al. [27] approach which models interferences.

This approach highlights three strong advantages. Firstly, it permits to consider an arbitrary number of senders. Secondly, it takes into consideration broadcast and unicast scenarios (adding further events, e.g. retransmissions, exponential backoff, collisions concerning ACK packets…). Thirdly, it matches with realistic traffic demands (not only infinite traffic demands).

The Qiu el al. model is based on an N-node Markov model (for capturing interactions among a number of senders) and on general and accurate sender and receiver models for both broadcast and unicast scenarios. It aims, among this, to estimate the goodput, the throughput and the loss rate between a pair of nodes with received signal strength (between every pair of nodes) and the traffic demand from each sender to each receiver as inputs.

To recapitulate our roadmap, as shown in figure 7, we try to master and to exploit many propositions. We use the model given in Zuniga & Krishnamachari [7, 8] approaches for the computation of packet reception rate and we project ton combine it with Qiu et al. contribution [27] to add interference impacts.

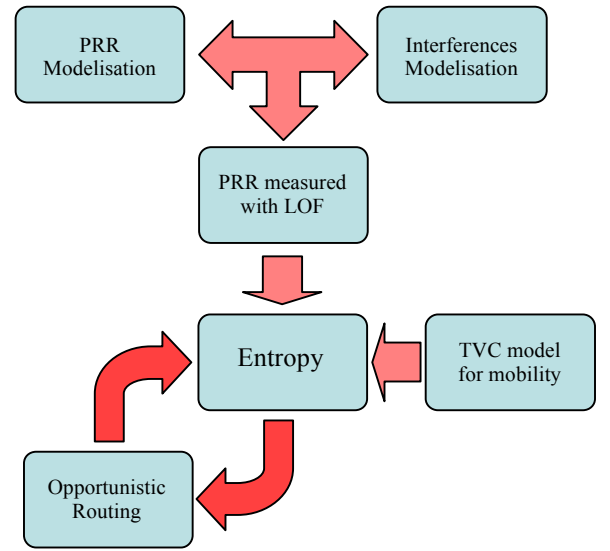

Figure 7. Roadmap of required models for the determination and the evaluation of entropy metric

The recourse to the sampling using the unicast traffic is accurate by referring to Zhang et al. works [10, 11] to compute efficiently the packet reception rate and avoid approximate beacons estimations.

We have also rise the challenge of inserting mobility. This task led us to adopt a mobility model that catches jointly the location preferences and time-variant behaviour.

All these propositions consider spatial and temporal aspects. The main objective behind the use of entropy metric is the aggregation of all these aspects in one measure, which will facilitate the routing process. The measure can give an idea about a neighbour, the neighbouring or even for an arbitrary subset of nodes located in the same geographical area. Then, the use of entropy is interesting for the opportunistic routing, giving that it simplify the identification of best routes.

## VI. Conclusion and Future Work

The wireless networks and especially wireless sensor network present interesting spatial and temporal properties. The recourse to the biased link sampling through data-driven link estimation helps to improve the routing process. Nevertheless, in order to fully grasp the true behaviour of a mobile environment we also need to take into account the spatiotemporal activity of the network. The measure of the entropy seems to be the most suited metric to track the spatiotemporal stability of networks. The coming challenge will be so to provide an expression that enables to compute this entropy, to identify the analytical parameters that will be taken into consideration and to explain the meaning of this entropy measure. Then, using simulations and testbed experimentations, with reliable mobility and accurate interference models, will be done in order to validate the pertinence of our proposal. We can also mention that we will relay on swarming to improve reliability and robustness of the network. Such recourse affords the possibility to catch more efficient information for a node from its neighbours, due to their close cooperation.